\documentclass[a4paper,11pt]{article}
\pdfoutput=1
\usepackage{jcappub} 
\usepackage{ragged2e}
\usepackage[T1]{fontenc} 
\usepackage{appendix}
\usepackage{graphicx}
\usepackage{dcolumn}
\usepackage{amssymb}
\usepackage{amsfonts}
\usepackage{amsbsy}
\usepackage{color}
\usepackage{rotating}
\usepackage[none]{hyphenat}
\usepackage[english]{babel}
\title{Status on Bidimensional Dark Energy Parameterizations Using SNe Ia JLA and BAO Datasets}

\author[a]{Celia Escamilla-Rivera,}
\affiliation[a]{Mesoamerican Centre for Theoretical Physics,
Universidad Aut\'onoma de Chiapas, Carretera Zapata Km. 4, Real del Bosque (Ter\'an), Tuxtla Guti\'errez 29040, Chiapas, M\'exico}

\emailAdd{cescamilla@mctp.mx}
    
\abstract{
Using current observations of forecast type Ia supernovae (SNe Ia) Joint Lightcurve Analysis (JLA) and baryon
acoustic oscillations (BAO), in this
paper we investigate six bidimensional dark energy parameterizations in order to explore which has
more constraining power.
Our results indicate that for
parameterizations that contain $z^2$-terms,
the tension ($\sigma$-distance) between these datasets seems to be reduced
and their behaviour is $<$1$\sigma$ compatible with the concordance model ($\Lambda$CDM).
Also, the results obtained by performing their
Bayesian evidence show a striking evidence in favour of the $\Lambda$CDM model, but only
one parameterization
can be distinguished by around $1\%$ from
the other models when the combination of datasets are considered.
}

\begin{document}
\maketitle
\flushbottom

\section{Introduction}
\label{sec:intro}

A highlight in observational cosmology is the origin of the
accelerated expansion of the universe.
The standard cosmological model that is consistent with current cosmological observations is the concordance model or $\Lambda$CDM.
According to this framework, the observed accelerating expansion is attributed to the
repulsive gravitational force of a cosmological constant $\Lambda$ with constant energy density $\rho$ and negative pressure
$p$. Despite its simplicity, this standard model has a couple of theoretical loopholes (e.g., the fine tuning and
coincidence problems) \cite{Weinberg:2000yb,Sahni:1999gb}, which had lead to alternative proposals that either modified the General
Relativity or consider a scenario with a dynamical dark energy. At~this point, dark energy can
be described by a parametrized equation of state (EoS) written in terms of the redshift, $w(z)$. Since its properties are
still under-researched, several proposals on dark energy parameterizations have been discussed in the
literature (see, e.g., \cite{Feng:2011zzo,Stefancic:2005sp,Jassal:2004ej,Wang:2008zh,Wang:2005yaa,
Barboza:2009ks,Pantazis:2016nky}).

The study of the constraints on the EoS parameter(s) has been done using observables such as: supernovae, baryon acoustic
oscillations (BAO), cosmic microwave background (CMB), weak lensing spectrum, etcetera. The importance of using these
compilations is due to the precision with which dark energy can be fathomed. Currently, some measurements such as the
Joint Lightcurve Analysis (JLA) from supernovae \cite{Ade:2015xua,Betoule:2014frx}, BOSS \cite{Busca:2012bu},
just to cite a few, point out a way to constrain these EoS parameters. These observations allow deviations
from the $\Lambda$CDM model, which are usually parametrized by a bidimensional form ($w_0$, $w_a$).

The aim of this paper is to study six bidimensional dark energy parameterizations, testing them with the SNe Ia and BAO data available and explore which one has more constraining power. \mbox{The organization} of this paper is as follows.
In Section \ref{sec:Mod_DE} we present how to model a parametrized dark energy via its EoS. In Section \ref{sec:bidim-DE} we
review six bidimensional dark energy parameterizations.
\mbox{The astrophysical} compilations to be use are described in Section \ref{sec:observational_data}. A description
of the Bayesian model selection is presented in Sections \ref{sec:bayesian} and \ref{sec:results}
we discuss our main results related to the \textit{tension}, \mbox{the Figure} of Merit and the Bayesian evidence for each dark energy parameterization. Our final comments are presented in Section \ref{sec:conclusions}.

\section{Modeling dark energy}
\label{sec:Mod_DE}

In order to achieve the observed cosmic acceleration, we require an energy density
with significant negative pressure at late times. This means that the ratio between the pressure and energy density is negative, i.e., $w(z)= P/\rho <0$. All reasonable fitting dark energy models are in agreement \mbox{at this point}.

We start with the Friedmann and Raychaudhuri equations for a spatially
flat universe
\begin{eqnarray} \label{hubble eq}
 E(z)^2&=&\left(\frac{H(z)}{H_0}\right)^2
 = \frac{8\pi G}{3}(\rho_m + \rho_{DE})  \left[\Omega_{0m} (1+z)^3 +\Omega_{0(DE)} f(z)\right] \quad
 \end{eqnarray}
and
\begin{eqnarray}
 \frac{\ddot{a}}{a} &=&
 -\frac{H^2}{2}\left[\Omega_m +\Omega_{DE} (1 + 3w)\right]
\end{eqnarray}
where $H(z)$ is the Hubble parameter, $G$ the gravitational constant and the subindex $0$ indicates the present-day values for the Hubble parameter and matter densities.
The energy density of the non-relativistic matter is $\rho_m (z)=\rho_{0m} (1+z)^3$. And the
dark energy density $\rho_{DE} (z)= \rho_{0(DE)} f(z)$, where
$f(z)= exp[\left[3\int^{z}_{0}{\frac{1+w(\tilde{z})}{1+\tilde{z}}}d\tilde{z}\right]]$.

We notice that modeling $w(z)$ can give directly a description
of the $E(z)^2$ function, as e.g., in the case of
quiessence models ($w = const.$) the solution of $f(z)$ is
$f(z)=(1+z)^{3(1+w)}$. If we consider the case of the cosmological
constant ($w=-1$) then $f =1$.\linebreak Other cases explore a dark energy density $\rho_{DE}$ with
varying and non-varying $w(z)$ (see, e.g., \cite{Stefancic:2005sp,Lazkoz:2005sp} and references therein).

\section{Bidimensional dark energy parameterizations}
\label{sec:bidim-DE}

In this section, we present the evolution of $E(z)^2$ for six bidimensional dark energy parameterizations most commonly used in the literature and we identify the parameters to be fitted using the current astrophysical data available.
\subsection{Lambda Cold Dark Matter-redshift parametrization ($\Lambda$CDM)}

Even though our first model has one independent parameter, $\Omega_{m}$, we shall take it into account
to compare with the bidimensional proposals. This model is given by:
\begin{eqnarray}\label{LCDM}
 E(z)^2&=&\Omega_m (1+z)^3 + (1-\Omega_m)
\end{eqnarray}
where we consider $w=-1$.

As it is well known in the literature, this standard model provides a good fit for a large number of observational data compilations
without addressing some important theoretical problems, such as the cosmic coincidence and the fine tuning of the $\Lambda$ value \cite{Peebles:2002gy}.

\subsection{Linear-redshift parametrization}

The dark energy EOS  for this case was presented in \cite{Huterer:2000mj,Weller:2001gf} and is given by:
\begin{equation}\label{linear}
\begin{aligned}
w(z) = w_0 - w_1 z
\end{aligned}
\end{equation}
which can be reduced to $\Lambda CDM$ model $(w(z)=w=-1)$ for $w_0 =-1$ and $w_1 =0$

Inserting Equation (\ref{linear}) into $f(z)$, we obtain
\begin{eqnarray}\label{linear p}
 E(z)^2 &=& \Omega_m (1+z)^3
 +(1-\Omega_m)(1+z)^{3(1+w_0 +w_1)}
\times e^{-3w_1 z}
\end{eqnarray}

However, this ansatz diverges at high redshift and consequently yields strong constraints on $w_1$ in
studies involving data at high redhisfts, e.g., when we use CMB data \cite{Wang:2007dg}.

\subsection{Chevallier-Polarski-Linder parametrization (CPL)}

A simple parameterization that shows interesting properties \cite{Chevallier:2000qy,Linder:2007wa} and, in particular, can be
represented by two parameters that exhibit the present value of the EoS  $w_0$ and its overall time evolution $w_1$ is the
CPL model, written as:
\begin{eqnarray}
 w (z)= w_0 +\left(\frac{z}{1+z}\right) w_1  \label{eq:cpl}
\end{eqnarray}
The evolution for this parameterization is given by:
\begin{eqnarray}\label{CPL}
 E(z)^2&=& \Omega_m (1+z)^{3}  +(1-\Omega_m)(1+z)^{3(1+w_0 +w_1)}
 \times e^{-\left(\frac{3w_1 z}{1+z}\right)}
\end{eqnarray}

\subsection{Barboza-Alcaniz parameterization (BA)}

Proposed in \cite{Barboza:2008rh}, this model brings a step forward in redshift regions where the CPL parameterization
cannot be extended to the entire history of the universe. Its functional form is given by:
\begin{eqnarray}
w(z)=w_0 + \frac{z(1+z)}{1+z^2} w_1
\end{eqnarray}
which is well-behaved at $z\rightarrow -1$. The evolution of this model can be written as:
\begin{eqnarray}\label{BA}
E(z)^2&=&\Omega_m (1+z)^{3}+(1-\Omega_m) (1+z)^{3(1+w_0)}
\times(1+z^2)^{3w_1/2}
 \end{eqnarray}

 \subsection{Low Correlation parameterization (LC)}

In \cite{Wang:2008zh} it was proposed a two parameter EoS for the dark energy component,
linear in the scale factor and given by:
\begin{eqnarray}\label{eq:w_wang}
w(z) &=& \frac{(-z +z_c) w_0 +  z (1+z_c) w_c}{(1+z)z_c}
\end{eqnarray}
where $w_{0} = w(z=0)$ and $w_{c} = w(z=z_{c}).$
The subindex $c$ is used to indicate the scale factor \mbox{(or redshift)} value for which the parameters $(w_{0},w_{c})$ are uncorrelated. This value depends on the different used data set. In this model was proposed to fix it at the value $z_{c}=0.5$ being this value sufficiently close to the current data value ($z_{c} \sim 0.3$) and thus arguing that the correlation between $(w_{0},w_{c})$ is relatively small.
With this value for $a_{c}$, the evolution now becomes
\begin{eqnarray}
E(z)^2 &=& \Omega_m (1+z)^3 + (1-\Omega_m) (1+z)^{3(1-2 w_0 +3 w_{0.5})}
\times e^{\left[\frac{9(w_0 -w_{0.5})z}{1+z}\right]} \;
\end{eqnarray}

The pivot $w_{0.5}$ is a conservative choice which achieved a low degree of correlation and provides
a simple expression.

\subsection{Jassal-Bagla-Padmanabhan parametrization (JBP)}

In \cite{Jassal:2004ej} another problem in CPL parametrization at high redshift $z$ was addressed. To alleviate this behaviour, the authors
proposed a new parametrization with the form
\begin{equation}\label{jassal}
\begin{aligned}
 w(z)= w_0 +\frac{z}{(1+z)^{2}} w_1
 \end{aligned}
\end{equation}
which can present a dark energy component with the same values at lower and higher redshifts, with~rapid variation at low $z$. Combining Equation (\ref{jassal}) and $f(z)$ we obtain
\begin{eqnarray}\label{JBP}
E(z)^2=\Omega_m (1+z)^{3} +(1-\Omega_m)(1+z)^{3(1+w_0)}e^{\frac{3w_1 z^2}{2(1+z)^2}}
\end{eqnarray}

\subsection{Wetterich-redshift parameterizations (WP)}

Another bidimensional parameterization was proposed in \cite{Wetterich:2004pv}, which include the possibility
that dark energy contributes to the total energy of the universe to some extent at an earlier epoch. Its form is given by:
\begin{equation}\label{wetterich}
\begin{aligned}
 w(z)= \frac{w_0}{{[1+w_1\ln{(1+z)}]}^{2}}
\end{aligned}
\end{equation}
where $w_1$ is called bending parameter and characterized the redshift where an approximately constant EoS turns over to a different behaviour.

Using Equation (\ref{wetterich}) in $f(z)$ we obtain the following evolution
\begin{equation}\label{WP}
\begin{aligned}
E(z)^2= \Omega_m (1+z)^{3}+(1-\Omega_m)(1+z)^{3\left[1+\frac{w_0}{1+w_1 \ln{(1+z)}}\right]}
\end{aligned}
\end{equation}

We may argue that the form of Equation (\ref{wetterich}) is not general enough and, in particular, not suitable for the description change of sign of $w(z)$. In fact, for typical models with early dark energy we expect $w(z)>0$ in the radiation era. However, this ansatz has some corrections when radiation becomes important \cite{Wetterich:2003qb}.

\section{Observational Data}
\label{sec:observational_data}

It is quite strongly stablished that dark energy domination began somewhat recently, and therefore low
redshift data, are precisely those best suited for its analysis. The two main astrophysical tools of such nature
are the standard candles (objects with well determined intrinsic luminosity) and standard rulers (objects with
well determinate comoving size). Such probes provide us with distance measures related to $H(z)$, and the
best so far representatives of those two classes are SNe Ia and BAO. Those are in fact low redshift datasets,
and much effort is begin done in those two observational contexts toward obtaining more and better measurements.

On one hand, SNe Ia are extremely rare astrophysical events, the modern and specifically planned strategies
of detection make it possible to observe and collect them up to relatively high redshift ($z\approx 2$). On the other hand, the
main techniques that rest on the BAO peaks detection in the galaxy power spectrum are promising standard rulers for
cosmology, potentially enabling precise measurements of the dark energy parameters with a minimum of systematic errors.

In the following lines we will describe the sources used for each astrophysical tools described above.


\subsection{Analysis using SNe Ia data}

To perform the cosmological test we will employ the most recent SNe Ia catalog available: \mbox{the JLA
\cite{Betoule:2014frx}}.
Its binned compilation shows the same trend as using the full catalog itself, for this reason we will use this reduced sample
which can be found in the above reference and explicitly \mbox{in \cite{Escamilla-Rivera:2016aca}}.
This dataset consist of $N_{\text{JLA}}=31$ events distributed over the redshift interval $0.01< z <1.3$. We remark that the
covariance matrix of the distance modulus $\mu$ used in the binned sample already estimated accounting
various statistical and systematic uncertainties. For further discussion see Section 5 in \cite{Betoule:2014frx, Conley:2011ku}.

To perform the statistical analysis of the SNe Ia we employ the distance modules of the \mbox{JLA sample}
\begin{eqnarray}
\mu(z_i, \mu_0) = 5 \log_{10}\left[(1+z)\int_{0}^{z}{d\tilde{z}E^{-1}(\tilde{z},\Omega_m;w_0,w_1)}\right] +\mu_0 \nonumber
\end{eqnarray}
where $(w_0, w_1)$ are the free parameters of the model.
and compute the best fits
by minimizing the~quantity
\begin{eqnarray}
\chi_{\text{SN}_{\text{JLA}}}^2
=\sum^{N_{\text{JLA}}}_{i=1}{\frac{\left[\mu(z_i ,\Omega_m ;\mu_0,
w_0,w_1)-\mu_{\text{obs}}(z_i)\right]^2}{\sigma^{2}_{\mu,i}}}
\end{eqnarray}
where the $\sigma^{2}_{\mu,i}$ are the measurements variances.

\subsection{Analysis using BAO data}
 
We also consider in our analysis the measurements of BAO observations
in the galaxy distribution. These observations can contribute important features by comparing the data of the
sound horizon today to the sound horizon at the time of recombination (extracted from the CMB anisotropy~data).
Commonly, the BAO distances are given as a combination of the angular scale and the \mbox{redshift separation}:
\begin{equation}
d_{z} \equiv \frac{r_{s}(z_{d})}{D_{V}(z)}, \quad \text{with} \quad  r_{s}(z_{d}) = \frac{c}{H_{0}} \int_{z_{d}}^{\infty}
\frac{c_{s}(z)}{E(z)} \mathrm{d}z
\end{equation}
where $r_{s}(z_{d})$ is the comoving sound horizon at the baryon dragging epoch,
$c$ the light velocity, $z_{d}$ is the
drag epoch redshift and $c^{2}_{s}= c^2/3[1+(3\Omega_{b0}/4\Omega_{\gamma 0})(1+z)^{-1}]$ is the sound speed
with $\Omega_{b0}$ and $\Omega_{\gamma 0}$ are the present values of baryon and photon parameters, respectively. By definition the dilation scale is
\begin{equation}
D_{V}(z,\Omega_m; w_0,w_1) = \left[ (1+z)^2 D_{A}^2 \frac{c
\, z}{H(z, \Omega_m; w_0,w_1)} \right]^{1/3}
\end{equation}
where $D_{A}$ is the angular diameter distance:
\begin{equation}
D_{A}(z,\Omega_m; w_0,w_1) = \frac{1}{1+z} \int_{0}^{z}
\frac{c \, \mathrm{d}\tilde{z}}{H(\tilde{z}, \Omega_m;w_0,w_1)} \;
\end{equation}

Through the comoving sound horizon, the distance ratio $d_{z}$ is
related to the expansion parameter $h$ (defined such that {$H \doteq 100 h$)} and the physical densities $\Omega_{m}$ and $\Omega_{b}$.

The BAO distances measurements employed in this paper are compilations of three surveys: $d_{z}(z=0.106)=0.336\pm 0.015$ from 6dFGS \cite{Beutler:2011hx},
$d_{z}(z=0.35)=0.1126\pm 0.0022$ from SDSS \cite{Anderson:2013zyy} and $d_{z}(z=0.57)=0.0726\pm 0.0007$ from BOSS CMASS \cite{Xu:2012hg}. Also, we consider three correlated measurements of $d_{z}(z=0.44)=0.073$, $d_z(z=0.6)=0.0726$ and $d_z(z=0.73)=0.0592$
from the WiggleZ survey \cite{Blake:2012pj}, with the inverse covariance matrix:
\begin{equation}
\mathbf{C^{-1}_{WiggleZ}}=\left(\begin{array}{ccc}
1040.3 & -807.5 & 336.8\\
-807.5 & 3720.3 & -1551.9 \\
336.8 & -1551.9 & 2914.9\\
\end{array} \right)\;
\end{equation}

The $\chi^2$ function for the BAO data can be defined as:
\begin{equation}\label{chibao}
\chi^2_{\mathrm{BAO}}(\boldsymbol{\theta}) =
\mathbf{X}^T_{\mathbf{BAO}}
\mathbf{C}^{-1}_{\mathbf{BAO}}
\mathbf{X}_{\mathbf{BAO}}
\end{equation}
where $\mathbf{X}_{\mathbf{BAO}}$ is given as
\begin{equation}
\mathbf{X_{BAO}}=\left(\begin{array}{c}
\frac{r_s (z_d)}{D_V (z,\Omega_m;w_0,w_1)})    - d_{z}(z)\\
\end{array} \right)\;
\end{equation}

Then, the total $\chi^2_{\mathrm{BAO}}$ is directly obtained by the sum of the individual quantity by using \mbox{Equation (\ref{chibao})} in:
$\chi^2_{\mathrm{BAO-total}}=\chi^2_{\mathrm{6dFGS}} +\chi^2_{\mathrm{SDSS}} +\chi^2_{\mathrm{BOSS CMASS}} +\chi^2_{\mathrm{WiggleZ}.}$

\section{Bayesian Evidence}
\label{sec:bayesian}

A Bayesian model selection is a methodology to describe the relationship between the cosmological model,
the astrophysical data and the prior information about the free parameters. \mbox{Using Bayes} theorem \cite{bayes-th}
we can updated the prior model probability to the posterior model probability. However, when we compare
models, the \textit{evidence} is used to evaluate the model's evolution using the data available. The evidence
is given by
\begin{eqnarray}\label{eq:bayes}
\mathcal{E} =\int{\mathcal{L}(\theta) P(\theta) d\theta}
\end{eqnarray}
where $\theta$ is the vector of free parameters, which in our analysis correspond to $(w_0, w_a)$ and $P(\theta)$
is the prior distribution of these parameters. Equation  (\ref{eq:bayes}) can be difficult to calculate due that
the integrations can consume to much time when the parametric space is large.
Nevertheless, even when several methods exist \cite{gregory,Trotta:2005ar}, in this work we applied a nested
sampling algorithm \cite{skilling} which has proven practicable in cosmology applications \cite{Liddle:2006kn}.

We compute the logarithm of the Bayes factor between two models $\mathcal{B}_{ij}=\mathcal{E}_{i}/\mathcal{E}_{j}$,
where the reference model ($\mathcal{E}_{i}$) with highest evidence is the $\Lambda$CDM model and impose a flat prior on $H_0$.
The interpretation scale known as Jeffreys's scale \cite{jeffreys},
is given as: if
$\ln{B_{ij}}<1$ there is not significant preference for the model with the highest evidence; if $1<\ln{B_{ij}}<2.5$ the
preference is substantial; if $2.5<\ln{B_{ij}}<5$ it is strong; if $\ln{B_{ij}}>5$ it is decisive.


\section{Results}
\label{sec:results}

Our main goal is to investigate the six bidimensional dark energy parameterizations
presented in Section \ref{sec:Mod_DE} and confronting them by using the SNe Ia JLA and BAO datasets in order to explore which has more constraining power and observe whether there is \textit{tension} between these two datasets, \mbox{which are} so far two of the most worthy tools to explore dark energy, and which are anticipated to play an even more preeminent
role in the future.

The process of considering dark energy constraints from the combination of SNe Ia JLA and BAO
datasets is relevant and useful, as is comparing the individual predictions drawn from each other. This
fact does not mean that we are going to completely avoid the use of the CMB analysis;  in particular, the selected priors
for $\Omega_m$ and $\Omega_b$ are obtained from a forecast of CMB observations with the Planck mission \cite{Ade:2015xua}.
The predicted best fits at 68\% confidence level are $\Omega_m =0.3089\pm 0.0062$ and $\Omega_b =0.0486\pm 0.0010$
with our choice for $H_0 =67.74\pm 0.46$ km s${}^{-1}$ Mpc${}^{-1}$.

\subsection{About the Likelihood and Tension}

We will employ the maximum likelihood method in order to determine the best fit values of the parameters
$w_0$ and $w_1$ for the six parameterizations described. The $\Lambda$CDM case can be set with $\Omega_m$ as an independent
parameter and compute its best fit. The total likelihood for joint data analysis is expressed as the sum of
each dataset, i.e.,
\begin{equation}\label{totalchi}
\chi_{\text{total}}^2 =\chi_{\text{SNe Ia}_{\text{JLA}}}^2
+\chi^2_{\text{BAO-total}}
\end{equation}

To compare results and test the tension among datasets, we compute the so called
$\sigma$-distance, \mbox{$d_{\sigma}$, i.e.,} the distance in units of $\sigma$
between the best fit points of the SNe Ia, BAO and the total compilation SNe Ia + BAO and the best fit points of each parameterization in comparison to the
$\Lambda$CDM model.
Following \cite{numerical}, the $\sigma$-distance is calculated by solving
\begin{equation}
1- \Gamma(1,\vert\Delta \chi_{\sigma}^2/2\vert)/\Gamma(1) = \mathrm{erf}(d_{\sigma}/\sqrt{2})
\end{equation}
where $\Gamma$ and $\mathrm{erf}$ are the Gamma and error function, respectively.
For homogeneity and consistency our `ruler' is in every case the total
$\chi^2$ function Equation (\ref{totalchi}), and our prescription is the \mbox{following \cite{EscamillaRivera:2011qb}}:
if we want to calculate the tension between SNe Ia and SNe Ia+BAO and the best fit parameters ([$w_0$,$w_1$])
then the previous $\Delta \chi_{\sigma}^2$ will be defined as $\chi_{tot}^2([w_0,w_1]_{\mathbf{SNe Ia+BAO}}) - \chi_{tot}^2([w_0,w_1]_{\mathbf{SNe Ia}})$; \mbox{other cases} follow this recipe.

\begin{figure}[H]
\centering
\includegraphics[width=1.0\textwidth,origin=c,angle=0]{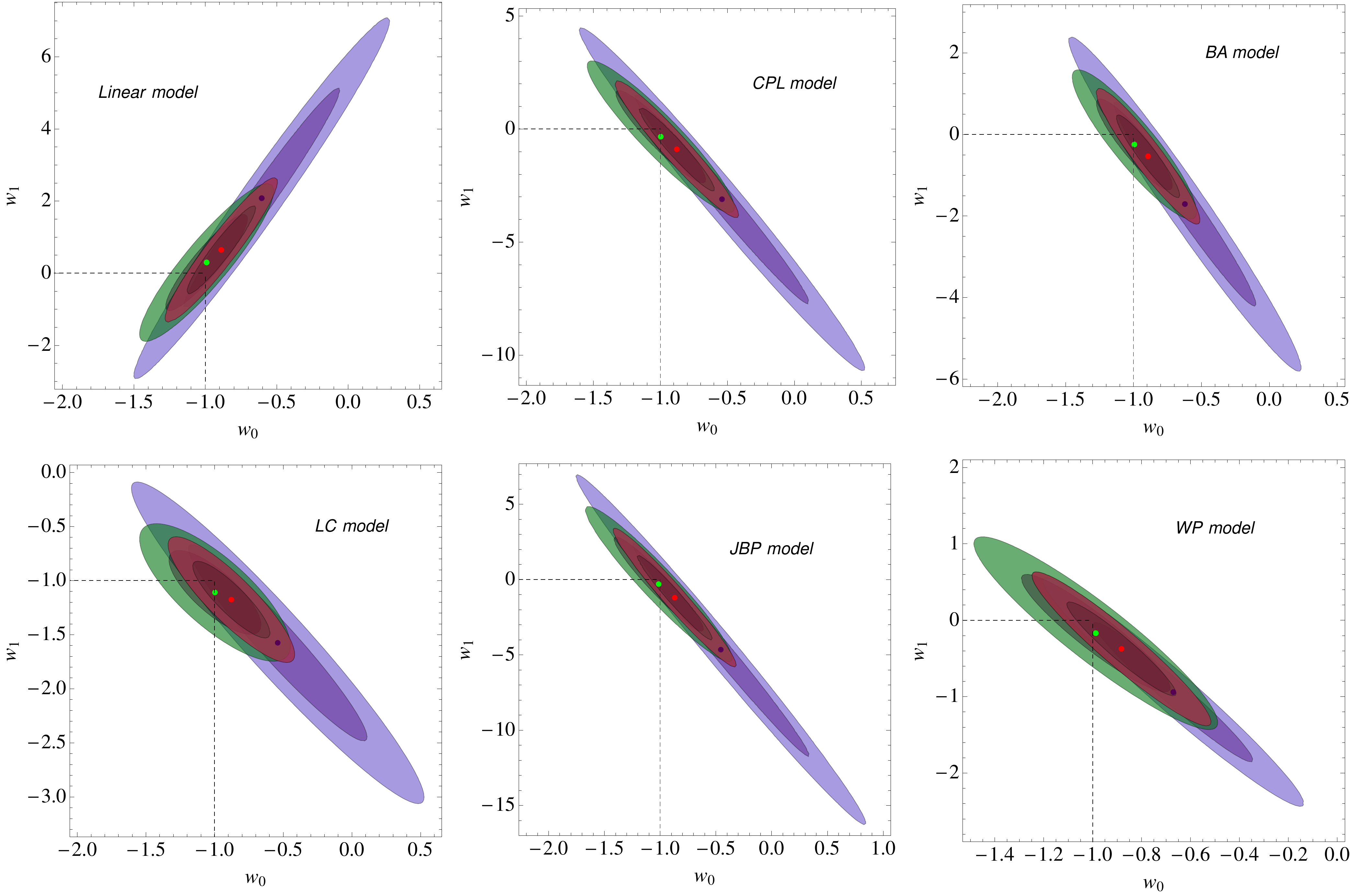}
\caption{\label{fig:cosmo_test4} 1 and 2$\sigma$ confidence contours for dark energy parameterizations.  Ia supernovae Joint Lightcurve Analysis (SNe Ia JLA) is represented by the green region, the baryon
acoustic oscillations (BAO) by the purple region and SNe Ia JLA+BAO by the red region. The best fits are indicated by the points for each sample, respectively. The point where the dashed line cross indicates the concordance model ($\Lambda$CDM).}\label{confidence_parameterizations}
\end{figure}

\begin{figure}[H]
\centering
\includegraphics[width=0.45\textwidth,origin=c,angle=0]{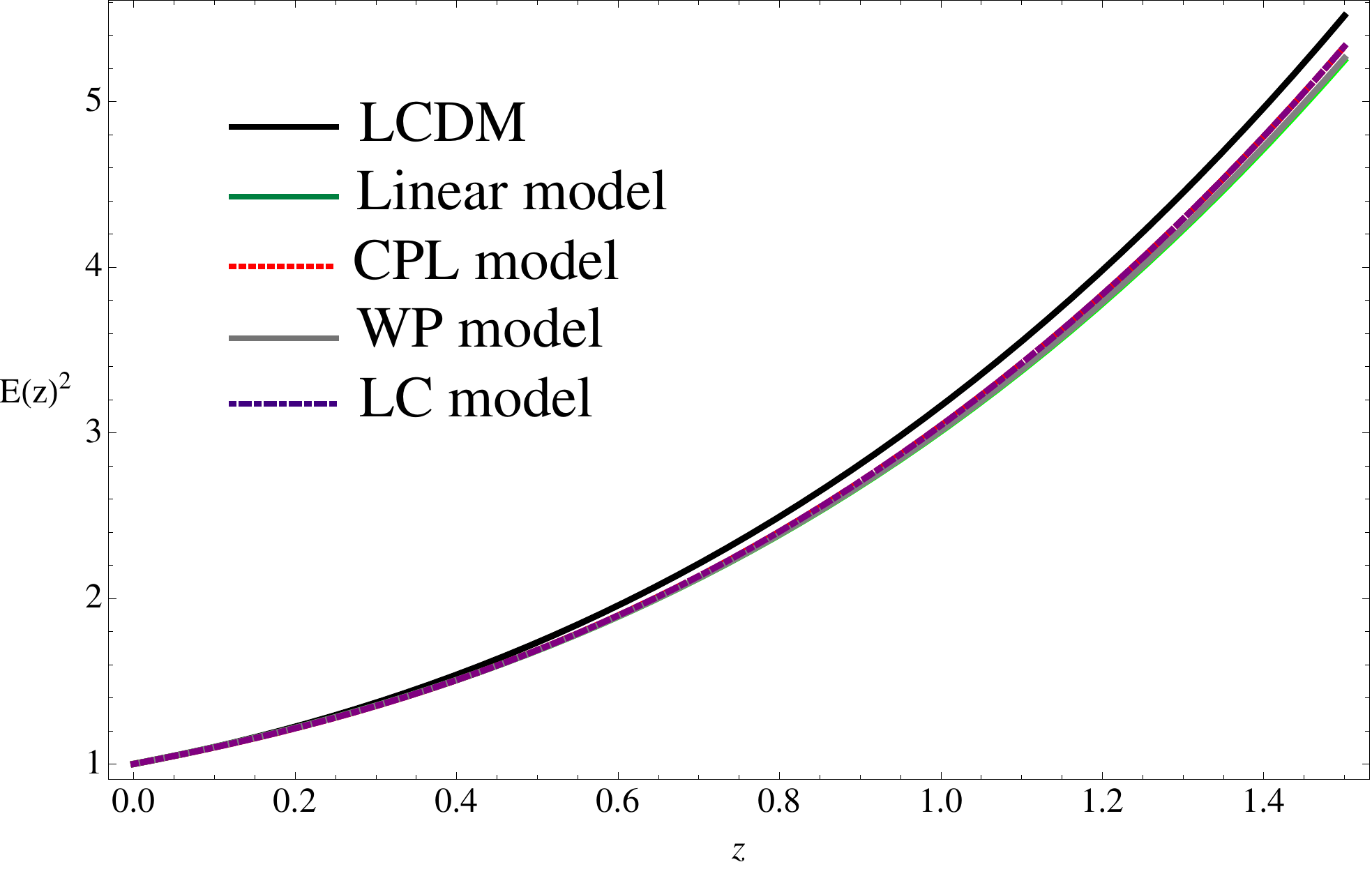}
\includegraphics[width=0.45\textwidth,origin=c,angle=0]{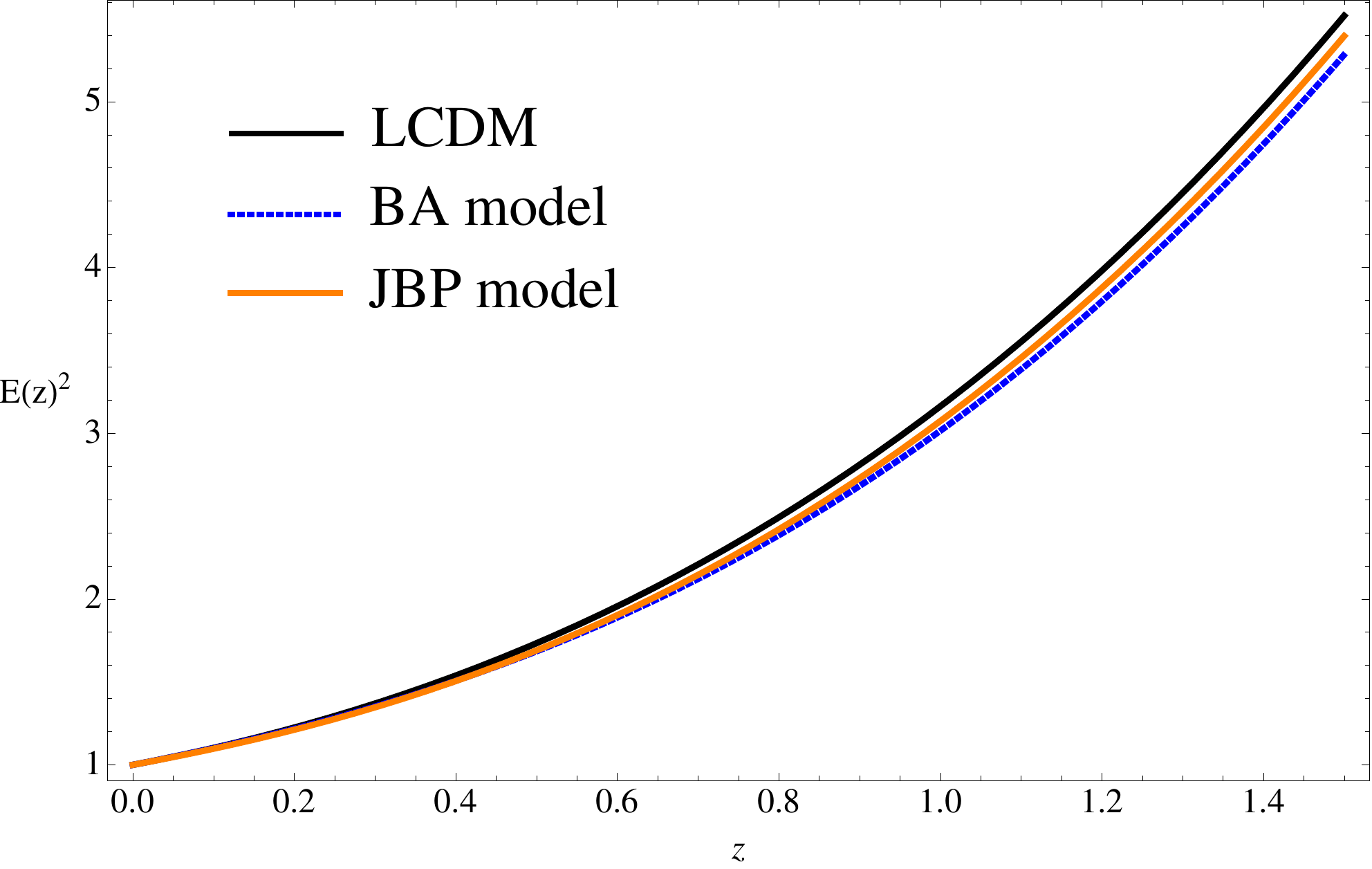}
\caption{$E(z)^2$ evolution function for each dark energy parameterizations.  We use the best fit obtained in each parameterizations with the SNe Ia JLA+BAO joined dataset. \textit{Left:} Evolution of Equation (\ref{LCDM}) and the bidimensional dark energy parameterizations Equations (\ref{linear p}), (\ref{CPL}) and (\ref{WP}). 
 \textit{Right:} Evolution of the Equations (\ref{LCDM}) and the bidimensional dark energy parameterizations  (\ref{BA})--(\ref{JBP}) (with $z^2$-terms in $w(z)$).
} \label{evolution_parameterizations}
\end{figure}

Looking at our results regarding the $\sigma$-distances in Tables \ref{table_bestfits} and  \ref{table_bestfits2} we can notice that the \textit{tension} between compilations
seems to be reduced when we use the parameterizations
that contain $z^2$-terms, as the BA and JBP models
(see Figures \ref{confidence_parameterizations} and \ref{evolution_parameterizations}).
Is important to address that this tension effect can change depending of the priors $\Omega_m$ and $\Omega_b$ as it was showed in  \cite {EscamillaRivera:2011qb},
but even with these changes, \mbox{the tension} remains reduced
for the BA and JBP parameterizations.


{\renewcommand{\tabcolsep}{-0.3mm}
\begin{table}[H]
\scalebox{0.80}[0.80]{
\begin{tabular}{|c|c|c|c|c|}
\hline {\bf Model}        &{\bf Parameterisation }    &$d_{\sigma}^{\Lambda CDM}$ & {\bf Best Fit Parameters using SNe Ia JLA data} \\
\hline LCDM & $ H^2(z)=H_0^2 [\Omega_{m}(1+z)^3 + (1- \Omega_{m})]  $    &$-$ &$\Omega_{m}=0.295\pm0.034$       \\
\hline Linear & $H^2 (z)=H_0^2 [\Omega_{m} (1+z)^3 + (1-\Omega_{m})(1+z)^{3(1+w_0+w_1)}e^{-3 w_1 z}]$ &
 $0.285$ &$w_0= -0.991\pm 0.036, \; w_1= 0.297\pm 0.779 $\\
\hline CPL & $H^2 (z)=H_0^2 [\Omega_{m} (1+z)^3 + (1-\Omega_{m})
(1+z)^{3(1+w_0+w_1)}$
&& \\ & $\times e^{-3 w_1 z\over{1+z}}] $ &$0.258$ &$w_0= -0.997\pm 0.049, \; w_1= -0.337\pm 1.822 $\\
\hline BA & $H^2 (z)=H_0^2 [\Omega_m (1+z)^{3}+(1-\Omega_m) (1+z)^{3(1+w_0)}$
&& \\ & $\times (1+z^2)^{3w_1 /2} $ 
&$0.243$ &$w_0= -0.993\pm 0.034, \; w_1= -0.245\pm 0.545 $\\
\hline LC & $H^2 (z)=H_0^2  [\Omega_m (1+z)^3 + (1-\Omega_m) (1+z)^{3(1-2 w_0 +3 w_{0.5})}$ && \\ & $ \times e^{\left[\frac{9(w_0 -w_{0.5})z}{1+z}\right]}] $ & $0.258$ &$w_0= -0.997\pm 0.049, \; w_{0.5}= -1.109\pm 0.066 $\\
\hline JBP & $H^2 (z)=H_0^2 [\Omega_{m} (1+z)^3 + (1-\Omega_{m})
(1+z)^{3(1+w_0)}$ && \\ & $\times e^{3 w_1 z^2\over{2(1+z)^2}}] $ &$0.236$ &$w_0= -1.013\pm 0.070 , \; w_1= -0.295\pm 4.306 $\\
\hline WP & $H^2 (z)=H_0^2 \left\{ \Omega_m (1+z)^{3}+(1-\Omega_m)(1+z)^{3\left[1+\frac{w_0}{1+w_1 \ln{(1+z)}}\right]}\right\}$  
&$0.278$ &$w_0= -0.987\pm 0.040 , \; w_1= -0.169\pm 0.258 $\\
\hline
\end{tabular}}
\caption{Dark energy parameterisations with best fits and $\sigma-$distances values using SNe Ia JLA data. 
} \label{table_bestfits}
\end{table}
}

{\renewcommand{\tabcolsep}{0.1mm}
{\renewcommand{\arraystretch}{1.5}
\begin{table}[H]
\begin{centering}
\scalebox{0.80}[0.80]{
\begin{tabular}{|c|c|c|c|c|c|}
\hline {\bf Model}        & {\bf Best Fit Parameters using BAO data}  &$d_{\sigma}^{\Lambda CDM}$ & {\bf Best Fit Parameters using SNe Ia JLA+BAO data}   &  $d_{\sigma_{Total}}^{\Lambda CDM}$ \\
\hline Linear & $w_0= -0.605\pm 0.130, \; w_1= 2.078\pm 4.063$ &$0.610$ &$w_0= -0.888\pm 0.025, \; w_1= 0.645\pm 0.650 $ &$0.380$\\
\hline CPL & $w_0= -0.540\pm 0.184, \; w_1= -3.105\pm 9.327 $ &$0.594$ &$w_0= -0.878\pm 0.034, \; w_1= -0.894\pm 1.487 $ & $0.323$ \\
\hline BA & $w_0= -0.621\pm 0.119, \; w_1= -1.707\pm 2.731$ &$0.600$ &$w_0= -0.892\pm 0.024, \; w_1= -0.535\pm 0.450$  & $0.316$\\
\hline LC & $w_0= -0.540\pm 0.184, \; w_{0.5}= -1.575\pm 0.359 $ &$0.594$ &$w_0= -0.878\pm 0.034, \; w_{0.5}= -1.175\pm 0.054 $ & $>1$\\
\hline JBP & $w_0= -0.456\pm 0.274, \; w_1= -4.653\pm 21.910$ &$0.569$ &$w_0= -0.869\pm 0.049 , \; w_1= -1.196\pm 3.441 $ & $0.257$\\
\hline WP & $w_0= -0.670\pm 0.046, \; w_1= -0.941\pm 0.363$ &$0.626$ &$w_0= -0.882\pm 0.022 , \; w_1=-0.375\pm 0.165$  & $0.386$\\
\hline
\end{tabular}
}
\vspace{0.2cm}
\caption{Dark energy parameterisations with best fits and $\sigma-$distances values using BAO and the combining samples.
}\label{table_bestfits2}
\end{centering}
\end{table}
}}

\subsection{About the Figure of Merit (FoM)}

In order to statistically compare our results, we compute, first, the Figure of Merit (FoM) as was proposed by the \textit{Dark Energy
Task Force} \cite{Albrecht:2009ct}, which is generally as the
$N$-dimensional volume enclosed by the confidence contours of the free parameters $(w_0,w_1)$ and
written as: FoM${}_{(w_0,w_1)}=1/\sqrt{\mathrm{det} \mathrm{Cov}(w_0,w_1)}$, with $ \mathrm{Cov}(w_0,w_1)$
the covariance matrix of the considered theoretical parameters. The FoMs for each dark energy
parameterizations are detailed in Table \ref{tab:fom}. From these values we notice that the FoM for
WP and LC parameterizations are better since they correspond to smaller error ellipse (see Figure \ref{confidence_parameterizations}). Also, we see that
BA parameterization shows a large parameter space volume in comparison to the JBP model.
\\

\subsection{About the Bayesian Evidence}

We estimate the \textit{evidence} using the algorithm discussed in \cite{Liddle:2006kn} and run it several times to obtain
a distribution of $\approx$100 values to reduce the statistical noise. Then we extract the best value to compute
the value of $\ln{B_{ij}}$, which is reported in Table \ref{tab:bayes} for each dark energy parameterization. As a result,  the $\ln{B_{ij}}$ values for each dark energy models lies in a region
in which $\Lambda$CDM is not discounted ($1<\ln{B_{ij}}<2.5$). These results show a striking evidence in favour of
the $\Lambda$CDM model. Moreover, BA parameterization display a $\ln{B_{ij}}$ of around 1\% larger than the other
parameterizations when SNe Ia+BAO is used.


\begin{table}[H]
\begin{center}
\caption{Values of the Figure of Merit for each parameterisation}
\scalebox{0.80}[0.80]{
\begin{tabular}{|c|c|c|c|c|c|c}
\hline
\multicolumn{1}{|c|}{$\mathrm{Model}$} & \multicolumn{3}{|c|}{FoM} \\
\hline \hline
$ $ &  SNe Ia JLA  &     BAO   & SNe Ia+BAO  \\
\hline
Linear & $14.203$ & $7.015$ & $23.657$ \\
CPL & $9.312$ & $4.631$ & $15.681$ \\
LC & $27.936$ & $13.893$ & $47.043$ \\
BA & $16.981$ & $8.555$ & $28.437$  \\
JBP & $6.076$ & $3.024$ & $10.337$ \\
WP & $26.208$ & $33.996$ & $53.677$ \\
\hline
\end{tabular}}
\label{tab:fom}
\end{center}
\end{table}

\begin{table}[H]
\begin{center}
\caption{Values of Bayes factor for each parameterisation.
}
\scalebox{0.80}[0.80]{
\begin{tabular}{|c|c|c|c|c|c|c}
\hline
\multicolumn{1}{|c|}{$\mathrm{Model}$} & \multicolumn{3}{|c|}{Bayes factor $\ln{B_{ij}}$} \\
\hline \hline
$ $ &  SNe Ia JLA   &     BAO   & SNe Ia JLA+BAO  \\
\hline
Linear & $1.904$ & $1.897$ & $1.857$ \\
CPL & $1.912$ & $1.903$ & $1.875$ \\
LC & $1.912$ & $1.903$ & $1.875$ \\
BA & $1.921$ & $1.921$ & $1.921$  \\
JBP & $1.918$ & $1.912$ & $1.854$ \\
WP & $1.906$ & $1.891$ & $1.855$ \\
\hline
\end{tabular}}
\label{tab:bayes_factor}
\end{center}
\end{table}

\section{Conclusions}
\label{sec:conclusions}
We have presented the study of six bidimensional dark energy parameterizations (Linear, CPL,
BA, LC, JBP and WP).
All of them were tested using
observations from SNe Ia JLA and BAO datasets, together with their combination.
Our results indicate that for
parameterizations with
$z^2$-terms in their $w(z)$-formulation (as BA and JBP models), the tension
between these datasets are reduced
and their behaviour is $<$1$\sigma$ compatible with $\Lambda$CDM.

Furthermore, for both parameterizations we have $w(z=0)=w_0$, but at high redshifts for BA $w(z\rightarrow \infty)=w_0 +w_1$
and for JBP $w(z\rightarrow \infty)=w_0$, this means that the JBP model can model a dark energy component which
has the same equation of state at the present epoch and at high redshift, while for the BA model we can rely on the results
only if $w_0 +w_1$ is below zero at the time of decoupling so that dark energy is not relevant for the physics of
recombination of the evolution of perturbations up to that epoch. Due to these behaviours we can consider that parameterizations
with $z^2$-terms are well-behaved and in better agreement with $\Lambda$CDM in comparison to other parameterizations where a \mbox{divergence is present}.

Also, the Bayes factor
shows striking evidence in favour of the $\Lambda$CDM model,
but the evidence for the concordance model is substantial with respect to the BA parameterization by around $1\%$ in comparison
to the other parameterization.
These results seems to be of interest since the bidimensional form of dark energy parameterizations are in better agreement with
$\Lambda$CDM,
wherever higher order parameterizations
can be developed.

We remark that these analyses were implemented to perform a complete treatment of selected $w(z)$
parameterizations along the lines of the study of contributions to the matter power spectra \cite{Escamilla-Rivera:2016aca}.

Complementary conclusions are that the use of statistical tools like Akaike Information Criterion (AIC) \cite{Liddle:2004nh}
can help us to discern between dark energy models that display different numbers of \mbox{free parameters.
}

\acknowledgments
Celia Escamilla-Rivera
is supported by CNPq Brazil Project 502454/2014-8 and
would like to thank David Polarski and George Pantazis
for their suggestions that improved the paper and Julio Fabris for his insights along these ideas.


\end{document}